\documentclass[manuscript,screen,nonacm]{acmart}

\AtBeginDocument{%
  }

\setcopyright{none}
\renewcommand\footnotetextcopyrightpermission[1]{}

\usepackage{mathtools} 
\usepackage{graphicx}
\usepackage{todonotes}
\usepackage{cuted} 
\usepackage{rotating}
	
\usepackage[nolist,nohyperlinks]{acronym}
\usepackage{pdflscape}
\usepackage{xcolor}
\usepackage{afterpage}
\usepackage{rotating}
\usepackage{draftwatermark}
\SetWatermarkText{DRAFT}
\SetWatermarkScale{0}

\begin{document}

\title{How Human-Centered Explainable AI Interface Are Designed and Evaluated: A Systematic Survey}

\author{Thu Nguyen}
\affiliation{%
  \institution{IT University of Copenhagen}
  \city{Copenhagen}
  \country{Denmark}}
\email{irng@itu.dk}

\author{Alessandro Canossa}
\affiliation{%
  \institution{Royal Danish Academy}
  \city{Copenhagen}
  \country{Denmark}}
\email{acan@kglakademi.dk}

\author{Jichen Zhu}
\affiliation{%
  \institution{IT University of Copenhagen}
  \city{Copenhagen}
  \country{Denmark}}
\email{jichen.zhu@gmail.com}


\begin{abstract}
Despite its technological breakthroughs, eXplainable Artificial Intelligence (XAI) research has limited success in producing the {\em effective explanations} needed by users. In order to improve XAI systems' usability, practical interpretability, and efficacy for real users, the emerging area of {\em Explainable Interfaces} (EIs) focuses on the user interface and user experience design aspects of XAI. This paper presents a systematic survey of 53 publications to identify current trends in human-XAI interaction and promising directions for EI design and development. This is among the first systematic survey of EI research.


\end{abstract}




\maketitle


\section{Introduction}

The research field of eXplainable AI (XAI) has emerged to make AI and machine learning (ML) more transparent and trustworthy to humans by opening the AI black-box and explaining its underlying operation\cite{gunning2017explainable}. This rapidly growing field has already made significant breakthroughs in {\em technical explainability}, producing established XAI algorithms such as  LIME~\cite{Ribeiro2016WhyClassifier}, DeepLIFT~\cite{Shrikumar2017LearningDifferences}, LRP~\cite{Binder2016Layer-WiseLayers}). 
In comparison, XAI research has limited success producing the {\em effective explanations} needed by users\cite{chromik_human-xai_2021,liao_human-centered_2022,xie2019interactive}. As a result, most explanations produced by XAI still lack usability, practical interpretability, and efficacy for real users~\cite{abdul2018trends, doshi2017towards,miller2019explanation, zhu2018explainable}. A recent study found that a significant group of users (over 30\%) were unable to understand the XAI explanations sufficiently well to use them even in relatively simple tasks 
~\cite{Narayanan2018HowExplanation}.

Recently there have been growing efforts, especially from the Human-Computer Interaction (HCI) community, to adopt more human-centered approaches and rigorous empirical evaluation methods~\cite{doshi2017towards,nguyen2022towards,liapis2022need} for XAI. For example, researchers draw from cognitive science theories of how people reason ~\cite{wang_designing_2019,villareale2022want} and from social sciences of how people explain ~\cite{Miller2017ExplainableSciences} and propose new paradigms of XAI that are grounded in existing understandings of human behavior. 

Another growth area is {\em explainable interface} (EI)\cite{mohseni_multidisciplinary_2021}, also referred to as explanation user interface\cite{chromik_i_2021} or explanation format\cite{chromik_human-xai_2021, mohseni_multidisciplinary_2021}, focusing on the user interface (UI) and user experience (UX) design aspects of XAI (e.g., structure, design, format, design, and content of the explanations). If the primary concern of XAI algorithms is to generate {\em what to explain}, EI research is about {\em how to explain} in a way that is effective for specific user groups. In other words, EI research is intrinsically user/human-centered. As a recently emerging area, EI research can shed light onto many fundamental questions about how to design XAI so that it can better meet the needs of real users. For example, there are currently no agreements over whether XAI should include all details of system logic\cite{Kulesza2015PrinciplesLearning, Kulesza2013TooModels} or only selective important information\cite{Herlocker2000ExplainingRecommendations, Schaffer2015GettingAnalysis}). Recently, Villareale et al. \cite{villareale2024GameXAI} explored the design possibility of EIs in the format of computer games.

While there are numerous review articles on the general trends in XAI\cite{adadi2018peeking,das2020opportunities,arrieta2020explainable,vilone2020explainable}, especially its technical development\cite{dovsilovic2018explainable,danilevsky2020survey,Supriyo2017,rojat2021explainable}, there has not been systematic efforts to map out the current state of EI. Even though the term EI only dates back to 2021, almost all XAI systems have an EI, whether it is explicitly designed or not. Furthermore, there has been related work in earlier research, most notably explanation facility in recommender system research\cite{tintarev2007explanations, nunes2017systematic}. 

Specially, we aim to identify current practices in current EI design approaches as well as promising directions that can further improve the usability, practical interpretability, and efficacy for real users of XAI. Towards these goals, we design our survey to focus on existing XAI research that involves human users, which we believe is a prerequisite of any human-centered XAI. Specifically, we seek to answer the following research questions: 


\begin{itemize}
    \item RQ1: \textit{How do researchers involve human participants in the design and development of XAI applications?}
    \item RQ2: \textit{How do XAI researchers design EIs?}
    \item RQ3: \textit{How do XAI researchers evaluate EIs?}
\end{itemize}

Using the Preferred reporting items for systematic reviews and meta-analyses (PRISMA) guideline\cite{moher2009preferred}, we included 53 publications for analysis.  
The remainder of the paper is organized as follows. We first present related review papers in XAI and describe our methods for the survey. Next, we analyze our results based on the above-mentioned three research questions. Finally, we discuss the implications of our findings.

\section{Related Work}
As the research area of XAI gain momentum, recently there have been numerous reviews about the general trends and opportunities of XAI\cite{adadi2018peeking,das2020opportunities,arrieta2020explainable,vilone2020explainable},  XAI for specific ML techniques (e.g., for supervised learning\cite{dovsilovic2018explainable}, for natural language process\cite{danilevsky2020survey}, for deep learning\cite{Supriyo2017}, and for time-series data\cite{rojat2021explainable}), and XAI in specific use domains (e.g., XAI for medical use\cite{tjoa2020survey,jimenez2020drug,payrovnaziri2020explainable}, for air-traffic management\cite{degas2022survey}). 

Most of the XAI review articles are technical in nature, while review articles on the design aspects of XAI research are relatively rare and recent. Amongst these design articles~\cite{mohseni_multidisciplinary_2021,tintarev2007explanations,oyindamola_williams_towards_2021}, summarizing existing design goals of XAI is a common approach, as design goals are a good way to capture research focus. 
For example, Mohensi et al.~\cite{mohseni_multidisciplinary_2021} investigate which design goals and evaluation metrics are used in XAI research. They found that in existing XAI research, for each of the three main user groups, there are common design goals associated with them: Novice Users (Algorithmic Transparency, User Trust and Reliance, Bias Mitigation, Privacy Awareness), Data Experts (Model Visualization
and Inspection, Model Tuning and Selection) and AI Experts (Model Interpretability, Model Debugging). They also classified five different types of evaluation measures: Mental Model, Usefulness and Satisfaction, User Trust
and Reliance, Human-AI Task Performance, and Computational Measures. 
Chromik and Butz~\cite{chromik_human-xai_2021} survey recommender systems, a sub-area of AI where early research on explainability has concentrated, and use Hornbæk and Oulasvirta's interaction types~\cite{hornbaek2017interaction} to classify different XAI design goals. For instance, "Dialogue" interaction is linked to the design goals of transparency and scrutability, while "Experience" is connected to satisfaction, trust, and persuasiveness.

Aiming to improve the practical effectiveness of XAI, a growing number of review papers attempt to bring insights from HCI/design research to this overwhelmingly of technical field. These reviews tend to focus on developing a deeper understanding of users and their needs. For instance, Suresh et al.~\cite{suresh2021beyond} propose that XAI research should differentiate the stakeholder expertise into knowledge and contexts, and stakeholder needs into long-term goals, shorter-term objectives, and immediate-term tasks. Based on an analysis of 58 publications, they find that their framework can help researchers to design more precise application-grounded evaluations. Similarly, Liao and Varshney~\cite{liao_human-centered_2022} expand the user groups of XAI to a broader range of stakeholders (Model developers, Business owners or administrators, Decision-makers, Impacted groups, and Regulatory bodies). Their survey found common disconnections 1) between technical XAI approaches and supporting users' end goals in usage contexts, 2) between assumptions underlying technical approaches to XAI and people's cognitive processes. They argue that technical choices of XAI algorithms should be driven by users' {\em explainability needs}. 

These existing works largely address the question of {\em WHAT} has been explained, leaving a gap about {\em HOW} to explain, especially in terms of how to design the explainable interfaces (EIs). Recent work~\cite{mohseni_multidisciplinary_2021} and~\cite{chromik_human-xai_2021} have call into attention the importance of explainable interfaces and started to develop design guidelines. Specifically, Mohensi et al.~\cite{mohseni_multidisciplinary_2021} proposed a design and evaluation framework where there is a separate layer dedicated to EI, including explanation format and interaction design. They call for future research to look into required features for both components of EI. 
Chromik and Butz's recent work~\cite{chromik_human-xai_2021} delves deeper into the interaction design of EI and examines different interaction styles in current XAI literature. While there is growing interest in EI design and development, there has not been a systematic review on this topic. To the best of our knowledge, this is among the first such work, and it will be complementary to the large body of algorithm-focused reviews of XAI research.

\section{Method}
\subsection{Dataset: Search Procedure and Inclusion Criteria}
We conducted a systematic survey to understand the current state of the art of human-centered XAI, especially in terms of the design and evaluation of explainable interfaces (EIs). We pay special attention to research that involves users since it is our belief that a user-centered design process is a prerequisite for human-centered XAI. 

Given that our focus on user participation in XAI and explainable interface is at the intersection of AI and HCI, we used ACM digital library (ACM DL) as the primary source for collecting publications. It includes relevant conferences such as CHI, Intelligent User Interface (IUI), and Designing Interactive Systems (DIS), where such work is regularly published. Since EI is a recently emerged research area\cite{mohseni_multidisciplinary_2021}, we acknowledge that it is possible that related work may be published in other venues, especially more technical AI conferences, or in arXiv. It is a limitation of our work.  

\begin{figure}[h!]
  \includegraphics[scale= 0.7, trim= 0cm 16cm 3cm 0cm]{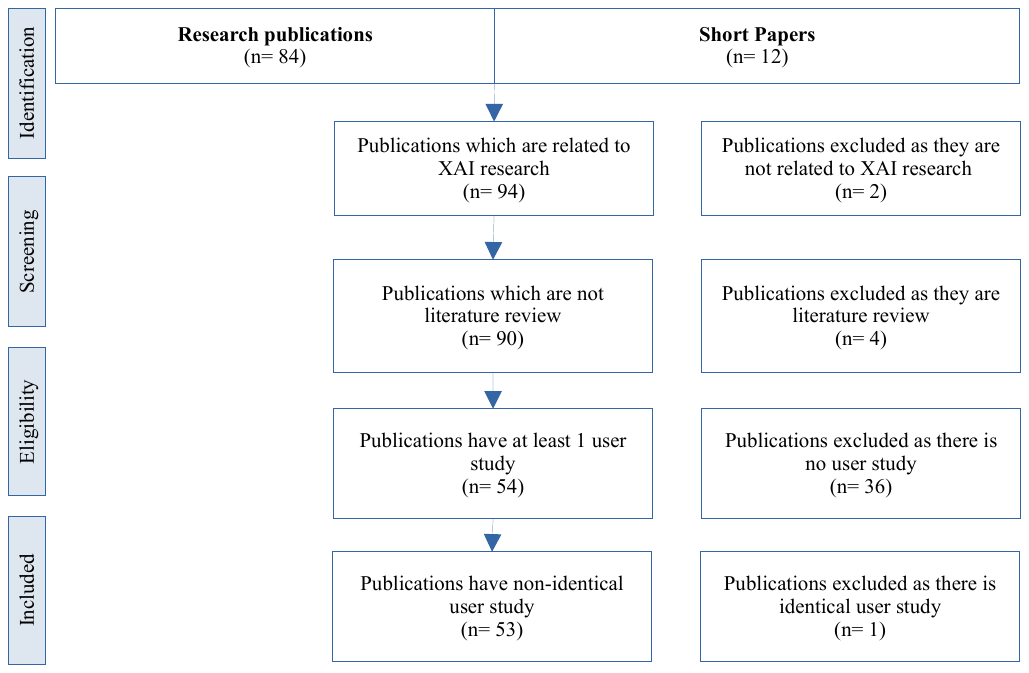}
  \caption{Publications selection process}
  \label{fig:PaperSelection}
\end{figure}

Our overall selection process is illustrated in Fig. \ref{fig:PaperSelection}. 


To cast a broad net, we query the ACM DL database to find scientific publications on XAI that mentioned "participants" or "users." Since \textit{"explainable interface"} is a new term first used by Mohensi et al. in 2021\cite{mohseni_multidisciplinary_2021}, we did not include it. Building on the query used by Chromik and Butz~\cite{chromik_human-xai_2021}, we also include the term \textit{"explanation facility"}, which the recommender systems research community uses to refer to explanations of the systems' recommendations. It is important to note that while the {\em term} of EI is new, almost all XAI has an interface, whether explicitly or implicitly designed, to convey the generated explanation to human participants.  
Finally, as XAI research became active in the last two decades, we include publications from January 2000 to May 2022. Below is our query:

\begin{center}
{\em "query": { Abstract:("XAI" OR "explainable AI" OR "explanation facility") AND Fulltext:("participant*" OR "user*") } 
"filter": { Article Type: Research Article, Publication Date: (01/01/2000 TO 05/31/2022), ACM Content: DL } \break
"filter": { Article Type: Short Paper, Publication Date: (01/01/2000 TO 05/31/2022), ACM Content: DL }}
\end{center}

From this query, a total of 96 publications (84 research articles and 12 short publications) are found for further screening. 
Next, we screen these publications to select the eligible ones.
Our {\em inclusion criteria} are 
\begin{enumerate}
    \item publications conduct research in the area of XAI (e.g., technical research on how to explain the decisions of AI, ML, recommendation systems, and other intelligent systems; or design/HCI research on how to design effective explanations), 
    \item publications that include at least one user study in the EI design process, including gathering user requirements, participatory design, and evaluation, except labeling training data sets, AND
    \item publications are not review papers.
\end{enumerate}

%



Our screening process is the following. We first read the abstract to identify publications to filter out those that clearly do not meet our inclusion criteria. Among the remaining ones, we go through the whole text to determine their eligibility. 

Several publications are excluded because they only mention user study superficially (e.g., referring to the authors' prior user study, or mentioning user study in planned future work) without reporting any empirical results. We also excluded publications with unsuccessful user studies where the study results are not reported (e.g.,~\cite{kou_mediating_2020}). In the case where there are multiple publications on the same user study, we only use the publication most relevant to EI. Out of the initial 96 publications, 53 passed our screening and are included in the data analysis.

\subsection{Data Analysis Methods}

\afterpage{%
    \clearpage%
    \thispagestyle{empty}%
    \begin{landscape}%
        \newgeometry{top=4.5cm,bottom=0cm,left=2cm,right=0cm, footskip=0cm, headsep=0cm}
        \noindent\begin{table}%
        \centering%
        \caption{Overview of 53 XAI publications that include human participation in the user studies}%
        \scalebox{.65}{%
        \begin{tabular}{  c c  c  c c  c c c  c c } %
                                            & Design     &information    &static/interactive   &interaction     &ISO steps       &participant group &  evaluation  &  metric type &  \\%
          Publications                           &   requirements     &  architecture   &          &  type  &    &         &   metrics   & & \\%
          \hline%
          Abdul et al. \cite{abdul_cogam_2020}     &F1, RF3        &sequential  &static &-- &Evaluate &general           &Cog1         &COG     & \\%
          Arrota et al. \cite{arrotta_dexar_2022}   &\textit{none}  &sequential &static &-- &Evaluate &domain           &Des5  &DES       &  \\%
          Bansal et al. \cite{bansal_does_2021}   &RF4     &sequential    &static                 &--    &Evaluate     &general              &Act3   &ACT       &   \\%
          Bove et al. \cite{bove_contextualization_2022}     &RF4, VH3, RF5     &hierarchical  &interactive      &instructing    &Evaluate   &general    &Cog1, Des5    &COG \& DES      &  \\%
          Bucinca et al. \cite{bucinca_proxy_2020}     &VH3         &sequential   &static &-- &Evaluate     &ML/AI, general      &Act3, Cog1, Des4, Des3      &COG, ACT \& DES        &  \\%
          Bucinca et al. \cite{bucinca_trust_2021}       &\textit{none}   &hierarchical  &interactive &instructing    &Evaluate      &general    &Des4, Des3, Cog1, Act3    &COG, ACT \& DES &    \\%
          Chromik et al.\cite{chromik_mind_2020}   &RF2, RF4   &sequential    &static     &--   &Analysis, Evaluate      &domain       &Act1, Des2        &ACT \& DES        & \\%
          Chromik et al. \cite{chromik_i_2021} &VH3, RF5     &matrix       &interactive     &instructing   &Evaluate     &general       &      &--        &\\%
          Das and Chernova \cite{das_leveraging_2020}     &RF2     &organic   &interactive      &exploring       &Evaluate     &general      &Act3         &ACT     &   \\%
          Das et al, \cite{das_explainable_2021}           &RF7      &sequential        &static     &--      &Evaluate      &general    &Act3    &ACT   & \\%
          David et al. \cite{ben_david_explainable_2021}             &VH1    &sequential     &interactive     &instructing       &Evaluate     &general        &Des4, Act4, Des5   &ACT \& DES   &\\%
          Dhanorkar et al. \cite{dhanorkar_who_2021}               &\textit{not designed}      &--         &static     &--        &Analysis        &ML/AI      &      &--  &    \\%
          Dodge et al. \cite{dodge_after-action_2021}     &RF4, VH1, VH2    &hierarchical  &interactive      &instructing       &Analysis, Evaluate     &domain   &     &--       &\\%
          Dodge et al. \cite{dodge_how_2022}           &VH3, RF5      &matrix     &interactive     &instructing        &Evaluate  &ML/AI, general       &Des4, Cog1, Act3       &COG, ACT \& DES  & \\%
          Donadello et al. \cite{donadello_explaining_2020}       &RF2     &sequential         &static     &--       &Evaluate  &domain    &Des2             &DES      & \\%
          Ehsan et al. \cite{ehsan_expanding_2021}        &RF8     &sequential      &static       &--      &Analysis, Produce, Evaluate     &ML/AI     &     &--   & \\%
          Flutura et al. \cite{flutura_RF5_2020}       &VH4, F1    &sequential      &static        &--      &Evaluate      &ML/AI  &Cog1     & COG  & \\%
          Ghai et al. \cite{ghai_explainable_2021} &\textit{none}      &sequential   &static     &--       &Evaluate     &ML/AI, general      &Des4, Des5, Cog1   &COG \& DES       & \\%
          Gorski and Ramakrishna \cite{gorski_explainable_2021}          &VH3   &sequential   &static     &--       &Analysis, Evaluate        &domain      &Cog1, Des2         &COG \& DES &  \\%
          Gou et al. \cite{guo_building_2022}  &F1, RF5      &organic        &interactive        &instructing  &Evaluate   &general      &Cog1, Des2, Act1, Act2, Des4, Des5     &COG, ACT  \& DES   &\\%
          Hadash et al. \cite{hadash_improving_2022}          &RF3         &sequential       &static     &--     &Evaluate &general       &Cog1  &COG    & \\%
          Hamon et al. \cite{hamon_impossible_2021}   &VH1, VH5, VH3          &sequential      &static        &--  &Produce     &ML/AI, general, domain    &  &--   &    \\%
          H.-Bocanegra and Ziegler \cite{hernandez-bocanegra_conversational_2021}    &RF2  &matrix          &interactive    &instructing \& conversing     &Analysis, Evaluate   &general     &Des1, Des4, Des2, Des5   &DES   &  \\%
          Hjorth \cite{hjorth_naturallanguageprocesing4all_2021}        &VH3, RF2     &organic      &interactive      &instructing \& manipulating      &Produce, Evaluate     &domain, general  &     &--   & \\%
          Jacobs et al. \cite{jacobs_ing_2021}       &RF5, RF6, VH3    &matrix      &interactive        &instructing      &Analysis, Produce, Evaluate      &domain   &Des2    &DES   & \\%
          Jesus et al. \cite{jesus_how_2021} &\textit{none}      &sequential   &static     &--       &Evaluate     &domain      &Act3, Des2  &ACT \& DES       & \\%
          Kaptein et al. \cite{kaptein_cloud-based_2022}          &RF2   &organic   &interactive     &instructing       &Evaluate        &general      &     &-- &  \\%
          Khanna et al. \cite{khanna_finding_2022}  &VH1      &hierarchical        &interactive        &manipulating  &Evaluate   &domain      &Act3     &ACT    &\\%
          Kim et al. \cite{kim_learn_2021}          &VH3, F1         &matrix       &interactive     &instructing     &Evaluate &ML/AI       &Act3, Act4, Cog1, Des4  &COG, ACT \& DES   & \\%
          Le et al. \cite{le_grace_2020}   &RF7         &sequential      &static        &--  &Evaluate     &general    &Des3, Des2, Cog1  &COG \& DES   &    \\%
          Lee et al. \cite{lee_exploratory_2020}               &\textit{not designed}   &--          &static    &--     &Produce   &domain     &  &--   &  \\%
          Liao et al. \cite{liao_questioning_2020}        &\textit{not designed}      &--      &static       &--      &Analysis     &UI UX     &     &--   & \\%
          Lima et al. \cite{lima_human_2021}       &\textit{not designed}     &--      &static        &--      &Analysis      &general   &     &--   & \\%
          Mai et al. \cite{mai_keeping_2020} &\textit{none}     &sequential   &static     &--       &Analysis, Evaluate     &domain      &Des2, Des3  &DES       & \\%
          Maltbie et al. \cite{maltbie_xai_2021}          &RF8   &sequential   &static    &--       &Evaluate        &domain      &Des2, Des3  &DES        & \\%
          Narkar et al. \cite{narkar_model_2021}  &VH3, VH4     &matrix        &interactive        &instructing  &Analysis, Evaluate   &ML/AI      &Des2    &DES   &\\%
          Nourani et al. \cite{nourani_anchoring_2021}          &RF4         &matrix       &interactive     &instructing     &Evaluate &general       &Act3, Des2, Cog1  &COG, ACT, DES   & \\%
          Panigutti et al. \cite{panigutti_understanding_2022}   &\textit{none}          &sequential      &static        &--  &Evaluate     &domain    &Des2  & DES   &    \\%
          Penney et al. \cite{penney_toward_2018}               &\textit{not designed}   &--          &static    &--     &Analysis   &domain     &  &--   &  \\%
          Polley et al. \cite{polley_towards_2021}  &RF7      &sequential        &interactive        &instructing  &Evaluate   &domain      &Des4, Des2, Cog1     &COG \& DES   &\\%
          Qian \cite{qian_evaluating_2022}          &RF5         &organic       &interactive     &instructing     &Evaluate &general       &Cog1, Cog2, Des3, Des3  &COG \& DES   & \\%
          Qiu et al. \cite{qiu_generating_2022}   &\textit{none}         &sequential      &static        &--  &Evaluate     &general    &Des2, Des5, Des4, Cog1  &COG \& DES   &    \\%
          Sevastjanova et al. \cite{sevastjanova_questioncomb_2021} &VH3, VH4,  VH1  &organic    &interactive    &instructing, manipulating \& exploring     &Evaluate   &domain     &Des2, Des4, Cog1, Act3  &COG, ACT \& DES   &  \\%
          Sklar and Azhar \cite{sklar_explanation_2018}  &RF2, RF5    &organic        &interactive        &conversing \& exploring  &Evaluate   &domain, general     &Act3, Des5     &ACT \& DES   &\\%
          Slijepcevic et al. \cite{slijepcevic_explaining_2022}          &VH4, VH3         &sequential       &static     &--     &Evaluate &domain       &Des2  &DES   & \\%
          Sovrano and Vitali \cite{sovrano_philosophy_2021}   &VH1, VH3,  VH5         &sequential      &interactive        &instructing  &Evaluate     &general    &Des2, Des5, Act3  &ACT \& DES    &    \\%
          Sun et al. \cite{sun_investigating_2022}               &\textit{not designed}   &--          &static    &--     &Analysis   &domain, ML/AI     &  &--   &  \\%
          Tabrez et al. \cite{tabrez_explanation-based_2019}          &RF6, RF8         &organic       &interactive     &exploring     &Evaluate &domain       &Des3  &DES   & \\%
          Wang et al. \cite{wang_ing_2019}   &VH3, VH4          &sequential     &static        &--  &Produce, Evaluate     &domain    &Cog1  &COG   &    \\%
          Wang and Yin \cite{wang_are_2021}               &\textit{none}   &sequential         &static    &--     &Evaluate   &general     &Cog1, Des4  &COG, DES   &  \\%
          Wang et al. \cite{wang_interpretable_2022}   &VH4, VH3, RF8         &matrix      &interactive        &instructing  &Evaluate     &general    &Des2, Des3  &DES   &    \\%
          Wolf \cite{wolf_explainability_2019}          &\textit{not designed}          &--       &--     &--     &Analysis &ML/AI       &  &--   & \\%
          Zhang and Lim \cite{zhang_towards_2022}               &RF8  &sequential          &interactive    &instructing     &Analysis, Evaluate   &general     &Des2, Act3 &ACT, DES   &  \\%
        \end{tabular}%
        }%
        
        \label{tab:overview53}%
        \end{table}%
    \end{landscape}%
    \clearpage%
}

We analyzed this data conducting thematic analysis~\cite{baxter2015understanding} and grounded theory~\cite{corbin2014basics}. 
Thematic analysis is used to categorize the \textit{required features and visual hierarchy} as well as \textit{evaluation metrics} and \textit{metric types}, these terms will be discussed later. Since we want to understand how EIs are currently designed and evaluated, we need to examine the properties that EIs possess. The properties of an EI, such as a graphical interface, can only be measured with qualitative method since they are concrete tangible features. Similarly, evaluation metrics which include different desired qualities and abilities that human can perform with the system, can also be accounted for with qualitative method. Thematic analysis is a qualitative method used to sort and arrange textual data into specific themes~\cite{baxter2015understanding}. Specifically, we collect the required features found in the selected publications and group similar features into the same category. For example, it is required for the explainable interface to have multiple explanations and to have multiple data input presented, these features are categorized into \textit{multiple instances} group. We use the online collaborative board platform {\em Miro} to conduct the thematic analysis. 

At the same time, we apply grounded theory to identify the abstract concepts that capture the presentation and navigation of the designed EIs. According to Corbin and Strauss~\cite{corbin2014basics}, a grounded theory approach generates theories from the data. To construct the theory, many iterations of data collection and theory construction are essential. Specifically, we conduct several  iterations of data collection and theory construction from different types of data related to the presentations of the EIs (e.g., format, interface elements, visualisation types, interaction type, UI type). In the \textit{Results} section we will list all the 8 properties that emerged from applying thematic analysis and grounded theory to our dataset, subdivided according to our three research questions.

\section{Results}
This section presents our findings based on our research questions (Table \ref{tab:overview53}) and a cluster analysis of the current trends in EI research. 

\subsection{RQ1: How do researchers involve human participants in the design and development of XAI applications?}

Based on our analysis, we find that two sub-questions are particularly useful to describe current practices: {\em at which stage} are human participants involved and {\em which type} of participants. 

In terms of {\bf the stage of involvement}, we analyzed our dataset of publications based on the widely used {\em ISO human-centered design process framework}~\cite{geis2016cpux}. In this standardized framework, after the \textit{Planning} activity, the iterative design process consists of 1) \textit{Analysis}: Understand and specify the context of use (e.g., User group profiles, task models, as-is scenarios, personas, user journey maps), 2) \textit{Specify} the user requirements (e.g., user needs definition, and related forms of understanding users), 
3) \textit{Produce} design solutions to meet user requirements, and 4) \textit{Evaluate} the designs against user requirements. Ideally, the development team should involve users and other stakeholders in all activities, especially in 1) and 4). This framework provides us with the first of the eight properties listed in \ref{tab:overview53}: \textit{Activities}. In table 1, we code the identified activities in our surveyed publications, except for Specify activity. Since Specify only indicates the results gathered from the Analysis activities and no specific actions found in this activity, we do not include Specify activity in our data analysis.

\begin{figure}[htb]
    \centering
    \includegraphics[width=.45\textwidth]{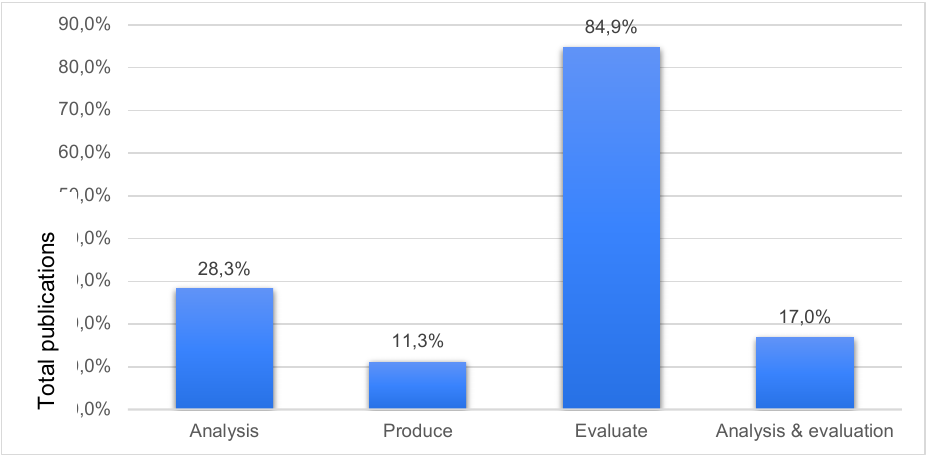}
    \includegraphics[width=.43\textwidth]{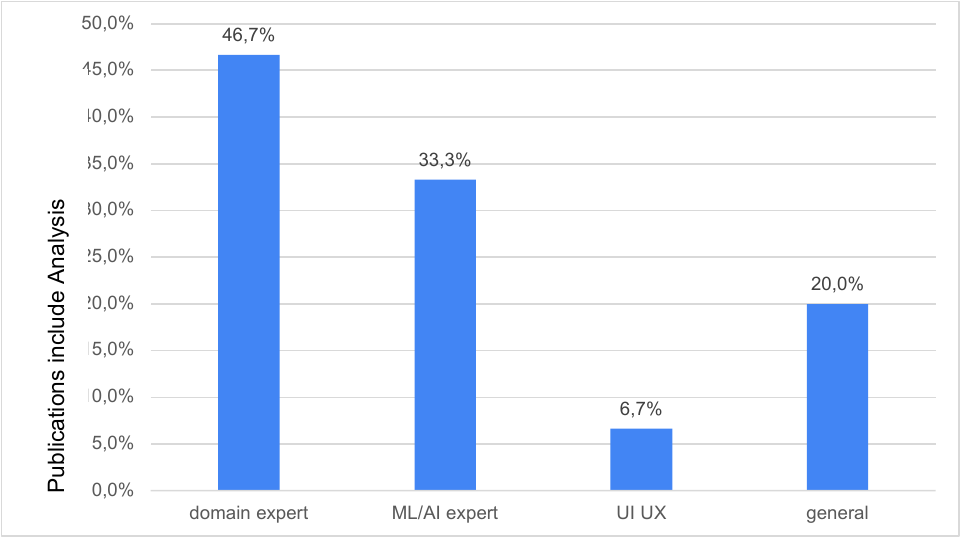}
    \caption{Left: Distribution percentage of the design activities that human participants are involved in (n=53). Right: Distribution percentage of the five main groups of human participants in the Analysis Activity (n=14).}
    \label{fig:UCDsteps}
\end{figure}

Fig. \ref{fig:UCDsteps} (Left) summarizes the human participants' involvement in different activities (counted by the number of publications). The overwhelming majority (84.9\%) of XAI publications involve humans to evaluate the XAI system, while 28.3\% in Analysis. The latter indicates that less than a third of the XAI research has conducted some form of user research before setting out to produce XAI solutions. Among all the publications, only 17\% involved humans in both Analysis and Evaluation, as the human-centered design process requires. While we acknowledge the differences between research and UX design projects, we believe that the ISO framework, especially at a high level, is useful to shed light on how current XAI research involves human participants. This is especially true because explanations are contextual and user group-specific and understanding users is an essential part of effective HCXAI. This is the second property, \textit{Participant groups}.

A main activity during the Analysis phase is defining user requirements, the results gathered from understanding users in the context of use that include user needs, and problems users have regarding the interactive system. Several publications explicitly gather user requirements for explanations (e.g.,~\cite{gorski_explainable_2021, hernandez-bocanegra_conversational_2021, jacobs_designing_2021, mai_keeping_2020, sun_investigating_2022}). Other publications have derived user requirements by understanding the challenges that users face in relation to XAI system (e.g.,~\cite{dhanorkar_who_2021, liao_questioning_2020}). 
We notice that all of these publications attempt to define user requirements (e.g. specific target user groups, user goals, user tasks) at the level of the XAI system in general. None reached the level of specificity of explainable interfaces, which results in a lot of ambiguity when the research team reach the Produce and Evaluation activities. 

Among the publications that gather user requirements, 60\% of them (17\% of the entire surveyed publications) evaluated the system using human participants. Some followed the established user centered design processes (e.g., the ISO framework). They first gather user requirements to drive their design, and then evaluate the prototype based on user requirements (e.g.,~\cite{chromik_mind_2020}). In contrast, some publications conducted user requirements gathering and prototype evaluation at the same time (e.g.,~\cite{gorski_explainable_2021, narkar_model_2021}). 
The latter case means that the prototype is not evaluated against user requirements.

In addition to defining user requirements, the ISO framework specifies other forms of understanding users in the Analysis activity. In our included publications, understanding users typically takes the form of 1) understanding how users make sense of the machine learning model (e.g.,~\cite{chromik_mind_2020, dodge_after-action_2021, ehsan_expanding_2021, gorski_explainable_2021, penney_toward_2018,wolf_explainability_2019, zhang_towards_2022}), 2) how they perceive AI (e.g.,~\cite{lima_human_2021}), or 3) how they perform XAI-related tasks such as comparing different ML models (e.g.,~\cite{narkar_model_2021}). All of these activities focuses on the users' cognitive processes, since the explanations produced by XAI has high information density to all user groups.




In terms of {\bf the type of human participants}, different user groups have different needs when it comes to explanations~\cite{liao_human-centered_2022}. Existing surveys have classified the stakeholder groups based on their ML expertise\cite{mohseni_multidisciplinary_2021} and their purposes of using XAI\cite{liao_human-centered_2022}. Similar to these surveys, we identified 3 stakeholders groups of domain experts, ML/AI experts, and UI/UX experts, and 1 general participant group. Fig. \ref{fig:UCDsteps} (Right) summarizes different groups involved in the Analysis activity. We find that 46.7\% of the surveyed publications involve domain experts such as caregivers~\cite{arrotta_dexar_2022}, lawyers~\cite{gorski_explainable_2021}, clinical experts~\cite{lee_exploratory_2020, jacobs_designing_2021, panigutti_doctor_2020, slijepcevic_explaining_2022, wang_designing_2019}. These XAI systems are typically designed to increase trust (e.g.,~\cite{chromik_mind_2020, panigutti_understanding_2022, polley_towards_2021}) and transparency~\cite{donadello_explaining_2020, hamon_impossible_2021,khanna_finding_2022} for domain experts. Another 33.3\% of our surveyed publications use participants with ML/AI expertise in their Analysis activity. This is typically for XAI systems designed for ML experts including tasks such as identifying existing problems~\cite{dhanorkar_who_2021}, ML experts' mental models~\cite{dodge_how_2022, ehsan_expanding_2021}), and improving the interpretability of the XAI system based on ML experts' insights~\cite{flutura_interactive_2020, hamon_impossible_2021}. The rest of the publications involve UI/UX practitioners and general participants, typically represented by the convenient sample of Amazon Mechanical Turk workers and students (represented as ``genernal'' in Fig. \ref{fig:UCDsteps} (Right)). UX/UI practitioners provide insights of user requirements for XAI system~\cite{liao_questioning_2020}. One main use of  general participants is to evaluate human-AI collaborating tasks (e.g.,~\cite{bansal_does_2021,bucinca_trust_2021, das_leveraging_2020, das_explainable_2021} and XAI system' desiderata (e.g., intepretability~\cite{bove_contextualization_2022, hadash_improving_2022}, trust~\cite{bucinca_trust_2021, guo_building_2022, ben_david_explainable_2021}, user satisfaction\cite{bove_contextualization_2022, ben_david_explainable_2021, guo_building_2022}). These are seemingly generic tasks without requiring specific user expertise. 

\subsection{RQ2: How are explainable interfaces currently designed?}



To answer RQ2, we analyze how the surveyed publications specify design goals and user experience, and what UI characteristics are represented in the final EI design. Compare to existing reviews that have examined the design goals of XAI overall\cite{mohseni_multidisciplinary_2021, oyindamola_williams_towards_2021}, our analysis is much more focused on the EI and how it facilitates the communication of technical explanations to human participants. 
Overall, 46 out of the total 53 publications (86.8\%) presented an explainable interface (EI) as part of an XAI system. 45 publications (84.9\%) designed and evaluated an EI, and 40 publications (75.5\%) reported design requirements explicitly. Again, we use the term EI broadly to capture the interface between the technical explanations and the users. Existing reviews have analyzed main EI design goals for specific user groups~\cite{mohseni_multidisciplinary_2021,chromik_human-xai_2021}. In this section, we attempt to understand, at a more concrete level, the information flow and the features that researchers decide that their EIs need to have. 



\begin{figure}
    \centering
    \includegraphics[width=.7\textwidth]{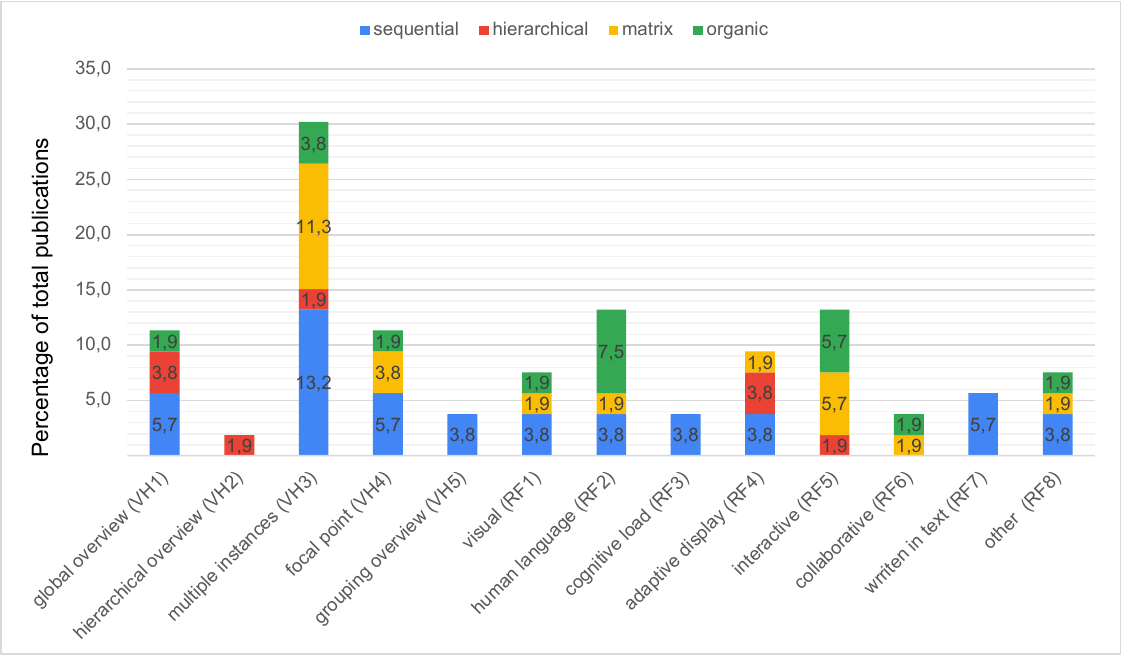}
    \caption{Design Requirements (including visual hierarchy and required features) correlated with the Selected Information Architecture in the Final EI Designs}
    \label{fig:requiredElem-Arch}
\end{figure}

\subsubsection{Design Requirements: Visual Hierarchy and Required Features} 



\par
Since the EIs in all surveyed publications use a graphical interface, we use the concept of \textit{visual hierarchy} to understand how researchers seek to organize explanation-related information in the design requirements. Visual hierarchy  
refers to ``the organization of the design elements on the page so that the eye is guided to consume each design element in the order of intended importance"~\cite{gordon_2021}. In our work, we analyze the 31 publications where the researchers explicitly report their design requirements  (e.g., explanation with natural human language~\cite{das_leveraging_2020, donadello_explaining_2020}, multiple explanations~\cite{gorski_explainable_2021, hamon_impossible_2021}, explanations with highlighting regions of interest~\cite{flutura_interactive_2020, narkar_model_2021}). We then use grounded theory~\cite{silverman2020qualitative} to derive five types of visual hierarchy: \textit{global overview} (VH1) ,\textit{hierarchical overview} (VH2), \textit{multiple instances} (VH3), \textit{focal points} (VH4), and \textit{grouping overview} (VH5). 

The left side of Fig. \ref{fig:requiredElem-Arch} shows the distribution of each visual hierarchy type over the number of publications. Note a publication can require multiple information hierarchy. For example, the required visual hierarchy from Sovrano et al.~\cite{sovrano_philosophy_2021} are global overview, multiple instance and grouping overview since they want to have many explanations that can be grouped and each of them provides global overview to confirm the theory that explanations can be understood by answering multiple answers from different questions.  The most common visual hierarchy requirements is for the EI to show multiple instances of explanations at the same time. These 16 publications (30.2\%) Multiple instances presents different factors of explanations or illustrates multiple examples of explanations in one view. For example, researchers report that their EI should show multiple explanations 
with different sets of input data for prediction~\cite{narkar_model_2021}, different pattern characteristics of the data~\cite{sevastjanova_questioncomb_2021}, or multiple explanations for comparison~\cite{chromik_i_2021, dodge_how_2022}. These publications choose multiple instances to help users to compare and contrast different instances and eventually help them build a general understanding of a machine learning model. 

The second common type of visual hierarchy requirement is global overview. 11.3\% the surveyed publications require explanations to be presented in a all-in-one style where different aspects of the explanations can be presented in the same view. Examples are that explanation should include multifaceted context~\cite{hamon_impossible_2021}, or be holistic to help users to locate bugs in explanations~\cite{khanna_finding_2022}, or provide overviews of different explanations~\cite{sovrano_philosophy_2021}.  

Another 11.3\% of the surveyed publications required focal point as the visual hierarchy of their EI design. They require regions of interest to be highlighted in the EI. An example is that explanations should highlight relevant input data that influence the prediction of the model~\cite{slijepcevic_explaining_2022}. 
In addition, we find 2 publications (3.8\%) require what we call grouping overview. In these publications, the researcher specify that the EI needs to present explanations in multiple clusters in the same view. Examples are that explanations should demonstrated in a group~\cite{hamon_impossible_2021}, or contextual information of explanation should be shown in group~\cite{sovrano_philosophy_2021}.
Finally, only 1 publication (1.9\%) require hierarchical overview to present explanations in a tree branching structure.
An example is that explanations represent different future states that an AI will be in based on the AI's possible sequences of actions~\cite{dodge_after-action_2021}.

\textit{Required features} are functionalities that the researchers claim that  their final EIs must have. These required features often have a large impact on the design space of the EI. For instance, the requirements of {\em being \textit{interactive}} and using \textit{human language} as output will highlight a chatbox interface as a design solution space. 
From our dataset of surveyed publications, 40 out of 53 publications (75.5\%) report explicit required features. 13 publications do not report required features (listed as "\textit{none}" in table 1).   

Using the Grounded Theory methods, we identified 8 main types of required features. The right side of Fig. \ref{fig:requiredElem-Arch} shows the distribution of each required feature. Relatively speaking, ``human language,'' being ``interactive,'' and ``adaptive display'' are the most common required features. 
Interactive interfaces, while technically more challenging to develop, are required by researchers to allow users to choose how to explore the EI (e.g.,~\cite{bove_contextualization_2022, chromik_mind_2020,qian_evaluating_2022}), to provide input to the system (e.g.,~\cite{guo_building_2022}). 
Notably, human language is required by the XAI researchers to explain the AI classifications (e.g.,~\cite{das_leveraging_2020}) or predictions in human understandable terms (e.g.,~\cite{chromik_mind_2020, donadello_explaining_2020,hjorth_naturallanguageprocesing4all_2021}). Some researchers chose human language so that users can converse with the XAI in the form of human dialog (e.g.,~\cite{hernandez-bocanegra_conversational_2021, sklar_explanation_2018}), 
or to make the system sound more intelligent~\cite{kaptein_cloud-based_2022}. The ``Other'' category include the least common features including audio feedback, recurrent explanation, and session recording.  
``User cognitive load'' related features are the least reported. Examples include the requirement to design the EI to reduce the users' cognitive load for visual information~\cite{abdul_cogam_2020} and semantic information (e.g.,~\cite{hadash_improving_2022}). This is the third property: \textit{required features and visual hierarchy}.

\begin{sidewaysfigure}
    \centering
    \vspace{15.5cm}
    \includegraphics[width=\columnwidth]{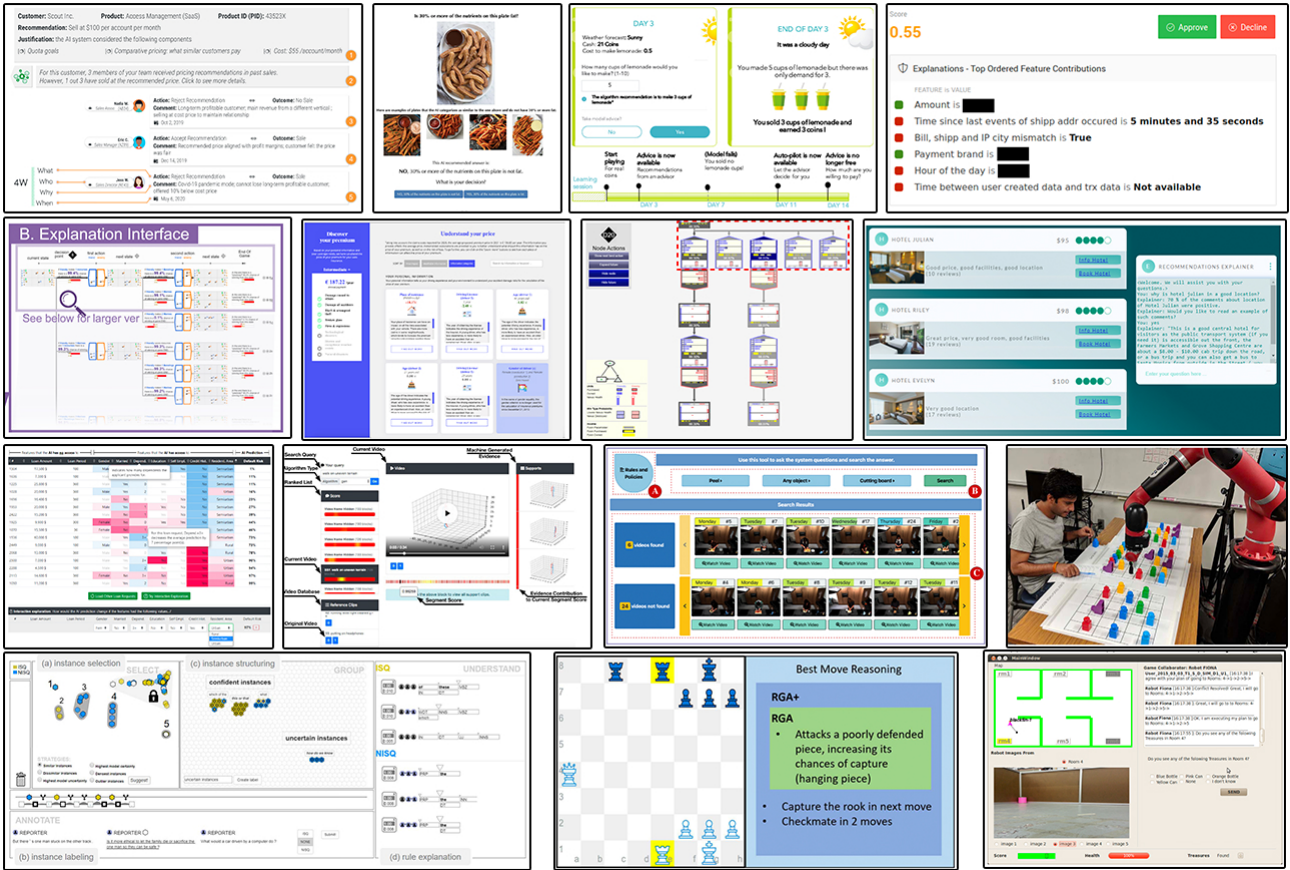}
    \caption{Explainable Interface Design Examples. Top Row (Left to Right):~\cite{ehsan_expanding_2021, bucinca_proxy_2020, ben_david_explainable_2021, jesus_how_2021}. Second Row:~\cite{khanna_finding_2022, bove_contextualization_2022, dodge_after-action_2021, hernandez-bocanegra_conversational_2021}. Third Row:~\cite{chromik_i_2021, kim_learn_2021, nourani_anchoring_2021, tabrez_explanation-based_2019}. Fourth Row:~\cite{sevastjanova_questioncomb_2021, das_leveraging_2020, sklar_explanation_2018}}
    \label{fig:collage}
\end{sidewaysfigure}

\subsubsection{Explainable Interface Design Outcomes: Information architecture, interactivity, and interaction type} After analyzing the reported design requirements of EIs, we examine the final design outcome of the EIs (examples can be seen in Fig. \ref{fig:collage}). 
Compared with the design requirements specified {\em prior to} the final design, we examine how the structure of the information from the system interface is carried out in the final design. Specifically, we classified the EI designs based on their \textit{information architecture} and whether they are \textit{interactive} or not. These are the fourth and fifth properties of our analysis. 

\begin{figure}
    \centering
    \includegraphics[width=.45\textwidth]{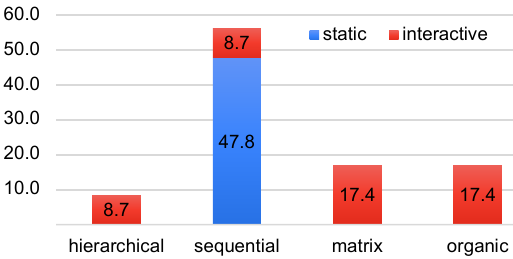}
    \includegraphics[width=.45\textwidth]{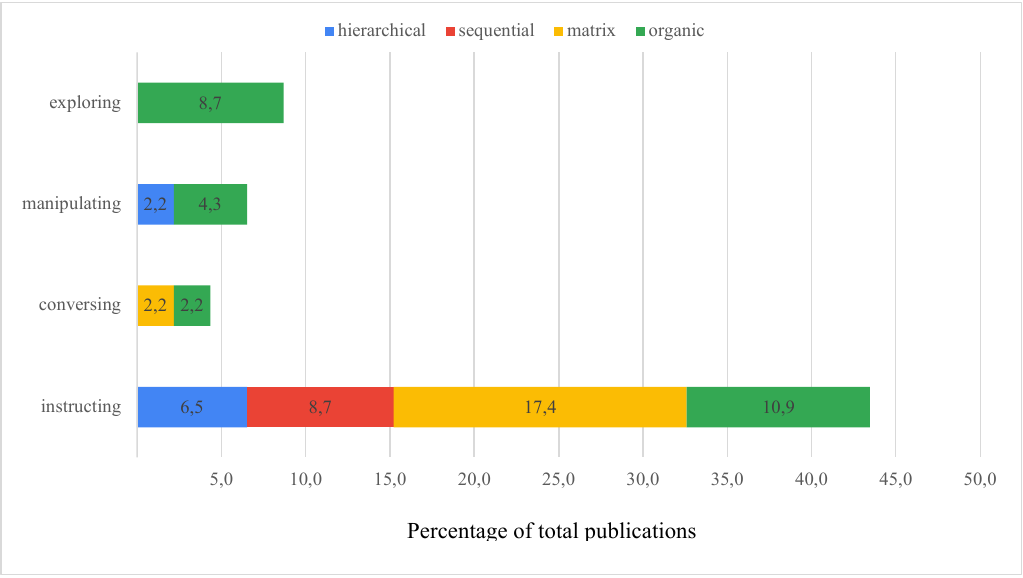}
    \caption{Left: designed information architecture. Right: Interaction types with designed information architecture}
    \label{fig:InfoArch_interactionTypes}
\end{figure}

\textit{Information architecture} indicates the way the content of a designed EI is organized. 
We adopt Garret's well-established taxonomy~\cite{garret2003elements}, which classifies UI into four primary types of information architecture. We then code all the EIs in our surveyed articles accordingly and our results are summarized in Fig. \ref{fig:InfoArch_interactionTypes} (Left). 




49.1\% of total surveyed publications that have the EI final designs included adopt the sequential information architecture, which presents information in a linear flow. In these EIs, users must go through explanations step by step in a pre-defined order. 
We believe that the main reason is that XAI explanations are typically highly technical. Sequential information architecture allows XAI researchers to control the order as well as the pace of how the explanations are presented to users. For example, Bucinca et al.~\cite{bucinca_proxy_2020} trained a ML model to predict the amount of fat content given images of food (Fig. \ref{fig:collage}, Second image Top Row). Their EI explains the prediction by showing to the users each of the food ingredients the ML model recognizes in the input image.  

Matrix structure and organic structure are less commonly used in final EI designs. Matrix structure ``allows user to move from node to node along two or more dimensions". A good example is Chromik et al.~\cite{chromik_i_2021}'s interactive EI that shows the data of 16 loan applicants (Fig. \ref{fig:collage}, First image Third Row). For each applicant, the EI shows the features the ML model uses, its decision for approving/denying the application, and other information (e.g., loan period). Using the spread sheet-like interface, this EI design allows users to sort the different features and even modify the values to see how the ML decision changes. By using the multiple explanations (16 data points) for comparison, the EI encourage users to understand the ML model by contrastive reasoning (comparing between the applicants) and counterfactual reasoning (e.g., what if this applicant had a different gender). Since the matrix structure is suitable to present different pieces of information at once, it is preferred to design explanations that require multiple instances, as seen in Fig. \ref{fig:requiredElem-Arch}. However, with organic structure design, the node of information evolves and depends on the actions that users perform with the system. For example, the chess game that users play with an agent from Das and Chernova~\cite{das_leveraging_2020} demonstrates this type of structure (Fig. \ref{fig:collage}, Third image Fourth Row). Specifically, since the moves from both the user and agent are completely unpredictable, each move a user makes will result in a different move that the agent will have, and vice versa. This type of interaction provides a free-form of interacting and exploring for users without a predefined structure.    



The least common information architecture (7.5\%) is hierarchical, which structures information in a tree-like branching structure. This is consistent with our above-mentioned analysis that hierarchical overviews is the least common design requirements. As shown in Fig. \ref{fig:requiredElem-Arch}, the one publication~\cite{dodge_after-action_2021} that requires hierarchical overview adopted hierarchical information architecture.

\textit{Interaction Types.}
Finally, Fig. \ref{fig:InfoArch_interactionTypes} summarizes the distribution of the interaction types~\cite{rogers2011interaction} of the final EI designs.
Rogers et al.'s framework identified four main types of interaction: 
1) {\em Instructing} is when user provide instructions, and commands to a system. Examples are typing commands, choosing options, and pressing buttons. 
2) {\em Conversing} is when user have a two-way communication conversation with a system. For example, user input format can be typing or speaking, and the system response are text or speech.
3) {\em Manipulating} is when "users interact with objects in a virtual or physical space by manipulating them". Examples are placing, opening, and picking.  
4) {\em Exploring} is when "users move through a virtual environment or a physical space". 


Our analysis shows that that Instructing is the predominant interaction type (37.7\%). All four information architecture structures are represented in the Instructing type. A main feature here is that users take control of how to approach the explanation, even though it typically takes limited forms such as clicking ``Next'' in a pre-defined sequence of explanations (e.g.,~\cite{arrotta_dexar_2022, bansal_does_2021, bucinca_proxy_2020, flutura_interactive_2020, ghai_explainable_2021, qiu_generating_2022})) . 
More substantive examples are that users can press the button to sort (e.g.,~\cite{bove_contextualization_2022}), filter (e.g.,~\cite{jacobs_designing_2021}), or even change the value of certain parameters (e.g.,~\cite{chromik_i_2021}). 
%
%
%
%
%
While conversing seems to be the most ``natural'' way for users to receive explanations, it is the least common (3.5\% of the surveyed publications, and 4.3\% of the designed EIs). 
In both of the two publications that use the Conversing interaction type~\cite{hernandez-bocanegra_conversational_2021, sklar_explanation_2018}, users interact with the XAI through a chat-bot. This is our sixth property. 


\subsection{RQ3: How are explainable interfaces currently evaluated?}



\paragraph{Evaluation metrics} We investigate which evaluation metrics are currently used to assess EI designs. We collect all the metrics from our surveyed publications, and apply thematic analysis to analyze patterns. Evaluation metric for system's desiderata includes \textit{transparency} (Des1), \textit{effectiveness} (Des2), \textit{efficiency} (Des3), \textit{user trust} (Des4), and \textit{user satisfaction} (Des5). User understanding of the system reflects on their cognitive ability \textit{to understand} (Cog1) and \textit{to learn} (Cog2) how machine learning model makes prediction. Since a user's understanding of the system influences the actions that they perform with the interface, their interaction can indicate their understanding of the system. In return, when users have limitation in interacting with the system, it influences their ability to understand and learn the system. Hence, it is necessary to examine user ability \textit{to control} (Act1), \textit{to respond} (Act2), and to collaborate: \textit{to accomplish task} (Act3) and \textit{to synchronize} (Act4) with the system. This is the seventh property identified. 

\begin{figure}
    \centering
    \includegraphics[width=0.75\textwidth]{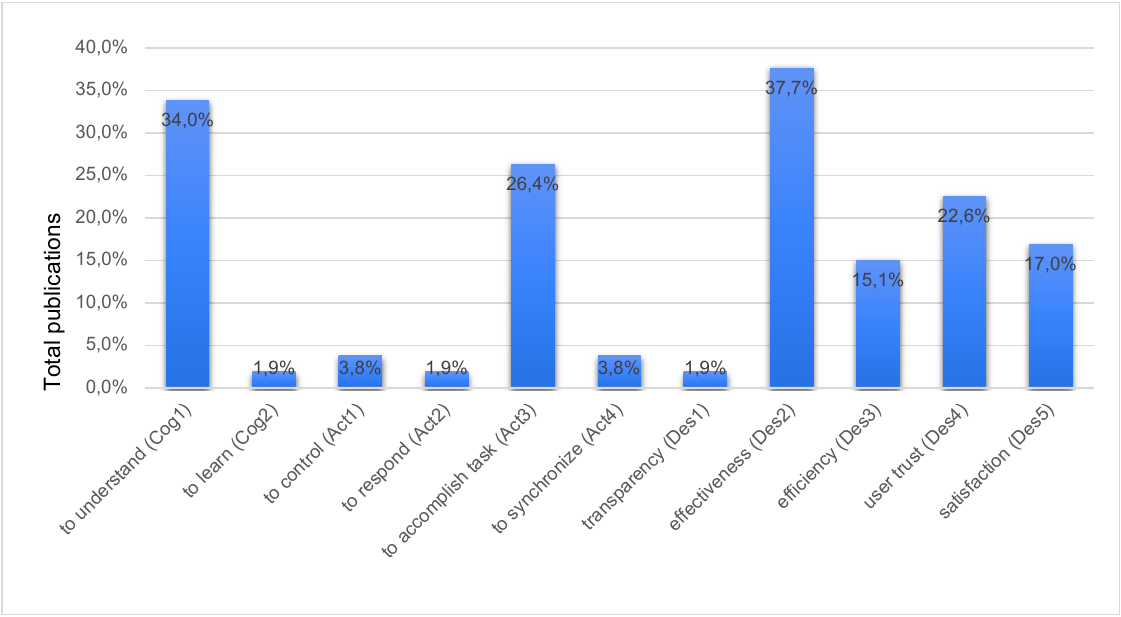}
    \caption{Specific evaluation metrics (including their types)}
    \label{fig:metrics2}
\end{figure}

\paragraph{Metric types}: The evaluation metrics used in our surveyed publications can be classified into system's \textit{desiderata} (DES) (60.4\% of surveyed publications), users' ability to perform \textit{cognitive} (COG) tasks (35.8\%), and users' ability to take certain \textit{actions} (ACT) (30.2\%) such as to learn, control, or respond. This is the eight and last property. It is worth noting that almost all publications evaluate the overall system where XAI is part of; no publication evaluated the EI alone except for only 1 publication from~\cite{bove_contextualization_2022} (where they evaluate the design principles of contextualization and allow exploration which are applied on the interface). The results in Fig. \ref{fig:metrics2} are from evaluations of the entire system. 

The popular area on the right of Fig. \ref{fig:metrics2} indicates that the majority of publications conduct evaluation for system's desiderata. System's effectiveness is the most common (37.7\%) evaluation metric for the XAI system. System's effectiveness is concered with accuracy (e.g.,~\cite{donadello_explaining_2020, maltbie_xai_2021}), helpfulness (e.g.,~\cite{jacobs_designing_2021, narkar_model_2021, sovrano_philosophy_2021}), and quality (e.g.,~\cite{mai_keeping_2020, panigutti_understanding_2022}). 

In terms of evaluating users ability to perform actions, the most common metric is examining how users accomplish tasks (26.4\%). Users' ability to accomplish tasks are measured with participants' object performance (e.g.,~\cite{bucinca_proxy_2020, das_leveraging_2020}), complete/error rate (e.g.,~\cite{das_leveraging_2020, das_explainable_2021, khanna_finding_2022, nourani_anchoring_2021, sklar_explanation_2018}), completion time (e.g.,~\cite{kim_explainable_2021, nourani_anchoring_2021, sovrano_philosophy_2021}). 

While performance task is a goal-driven metric, other action-driven metrics area not often measured. There are a few publications that evaluate users' ability to control (3.8\%) (e.g., a sense of control users have~\cite{chromik_mind_2020, guo_building_2022}), and ability to respond (1.9\%).  



\begin{figure}
    \centering
    \includegraphics[width=0.95\textwidth]{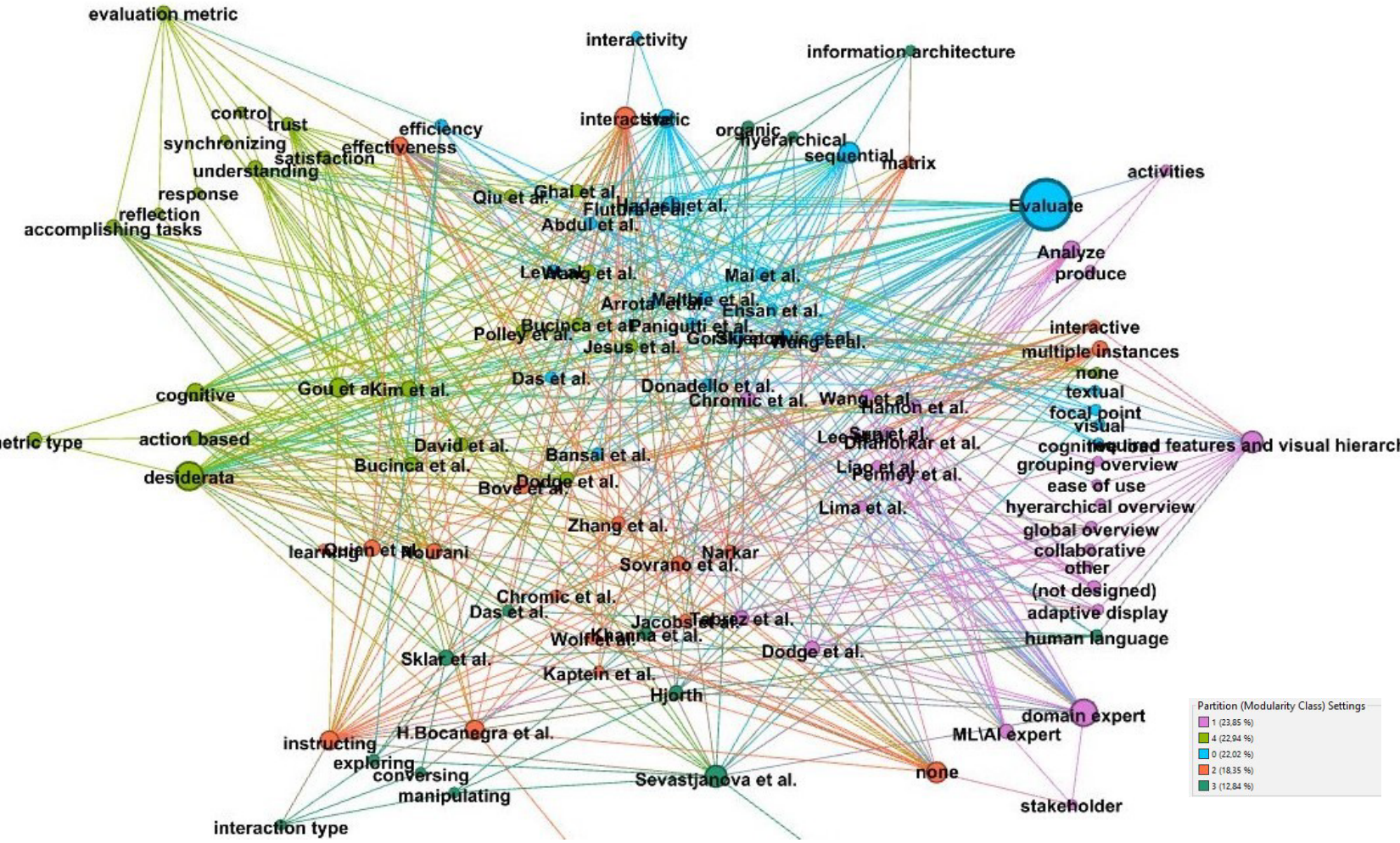}
    \caption{Overall Cluster Analysis of Literature}
    \label{fig:cluster_all}
\end{figure}

\subsection{Cluster analysis of literature}
In order to unfold more complex multidimensional relations within the XAI field, we decided to cluster all the surveyed publications (see Fig. \ref{fig:cluster_all}). The network graph was created in Gephy (https://gephi.org/), an open-source network analysis and visualization package, using the modularity measure. The intent was to identify clusters of publications with a similar configuration of the 8 properties listed in Table \ref{tab:overview53} (\textit{Participant group, Activities, Required Features and Visual Hierarchy, Interaction Type, Interactivity, Information Architecture, Evaluation Metric, Metric Type}). Modularity is a measure used in Social Network Analysis and is defined as the fraction of the edges that fall within the given groups minus the expected fraction if edges were distributed at random; while communities are defined as having dense connections between the nodes within modules but sparse connections between nodes in different modules. Modularity was preferred to traditional clustering algorithms such as K-means because of three reasons: a) modularity analysis in Gephy allows visualizing the communities detected on a network graph, this feature is highly prized as it facilitates human interpretation and understanding of the nature of the resulting clusters, b) modularity performs well with highly dimensional but sparse data, while both k-means and k-mode struggle with that, and c) modularity is a-parametric and does not require assumptions such as stating a priori the number of desired clusters, as it is necessary with k-means. Gephy implements modularity analysis with the Louvain method. 
In the network graph shown in Fig. \ref{fig:cluster_all}, the colors represent the 5 communities (or clusters) identified. Cluster 0 (light blue) captures 22.02\% of the publications examined, cluster 1 (pink) covers 23,85\%, cluster 2 (orange) includes 18,35\%, cluster 3 (dark green) encloses 12,84\% and cluster 4 (light green) catches the remaining 22,94\%. At the center, all 53 publications are represented, while on the periphery of the graph the 8 properties with their relative categories are used to define the publications. As Fig. \ref{fig:cluster_all}, most publications involved humans in evaluation.  Below, as an example, we will analyze the largest clusters 0 and 1.

\begin{figure}
    \centering
    \includegraphics[width=0.47\textwidth]{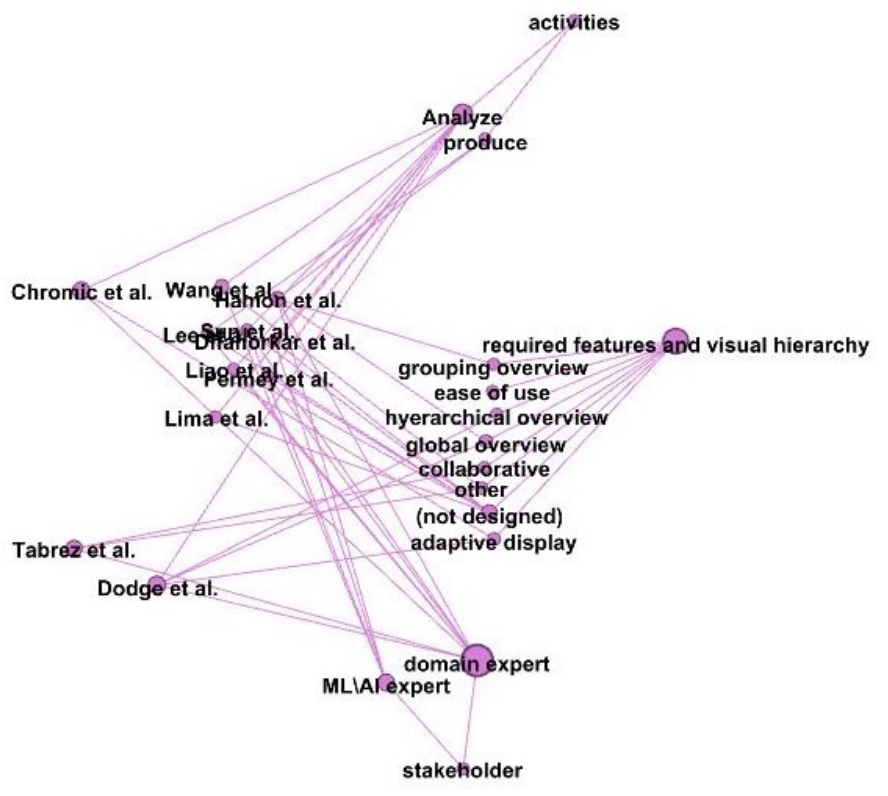}
        \includegraphics[width=0.47\textwidth]{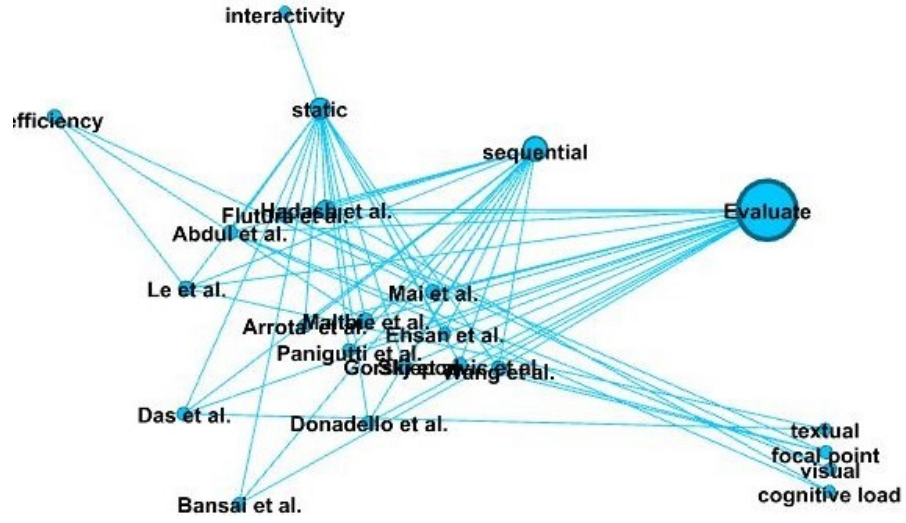}
    \caption{Left: Detailed View of Cluster 1. Right: Detailed View of Cluster 0}
    \label{fig:cluster_0_1}
\end{figure}

\subsubsection{Cluster 1.}
As shown in Fig. \ref{fig:cluster_0_1} (Left) this cluster is defined by \textit{Domain Experts} as stakeholders sometimes paired by \textit{ML\textbackslash AI experts} as well as \textit{Analysis} and \textit{Produce} activities.

This pattern is indicative of a need to understand how humans make sense of XAI system in the context of games. For example, Dodge et al.~\cite{dodge_after-action_2021} attempt to understand human reasoning processes when performing an assessment of the system's decisions in a real-time strategy game. Also, domain experts' sense-making can confirm whether the applied cognitive theories in XAI system are effective. For example, Information Foraging Theory relates to the understanding of how humans search for information, this theory is adopted in the context of a real-time strategy game to understand how domain experts (players) understand and seek explanations for the actions of the system~\cite{penney_toward_2018}. Similarly, Tabrez el al.~\cite{tabrez_explanation-based_2019} evaluate the cognitive theory of how humans cope with misunderstandings of a system's behavior by introducing rewards and explanations. We assume that since domain experts are already familiar with the studied domain and thus require less training to interact with the system, they require less effort in making sense of the domain. In contrast, if they are unfamiliar with the domain, they might have to put more effort into making sense of the new domain. Hence, the sense-making process for domain experts is less likely to be affected by the unfamiliar domain. We suspect that this could be the reason why domain experts are preferred when evaluating cognitive theories.

This cluster is also defined by a pattern of activities that exclude \textit{Evaluate} but focuses on \textit{Analize} and \textit{Produce}, We hypothesize that domain experts are more involved in the \textit{Produce} activity because the domains under examination are higher stake domains such as clinical and medical. Specifically, intensive care unit (ICU) clinicians are part of the co-design sessions to provide their diagnostic reasoning process and the usage of explanation features as they interact with the interface~\cite{wang_designing_2019}. Similarly, ICU medical staff were in the loop to validate the decision that the AI system makes on the diagnosis of lung cancer~\cite{hamon_impossible_2021}. Also, therapists annotate the exercise dataset from stroke and healthy subjects~\cite{lee_exploratory_2020}. Since clinical and medical domains have higher stakes for users, it is essential to include domain experts while building the system.

\subsubsection{Cluster 0.}
As shown in Fig. \ref{fig:cluster_0_1} (Right) this cluster is defined by \textit{Evaluate} activity,  \textit{Sequential} information architecture and \textit{Static} interactivity.

Publications that involve participants to evaluate the XAI systems tend to adopt 'black-box' machine learning models and apply agnostic approaches to explain how the model works. Since the main problems of models such as convolution neural networks (e.g.,~\cite{gorski_explainable_2021}), deep neural networks (e.g.,~\cite{arrotta_dexar_2022}), neural networks (e.g.,~\cite{flutura_interactive_2020, le_grace_2020}) is low interpretability, XAI researchers prioritize to evaluate the generated explanation with the interpretability criteria. 

Furthermore, since the textual format communicates clearly to users, XAI researchers adopt textual explanations to increase the interpretability level of such black-box models. Specifically, they design user studies with text classifications (e.g.,~\cite{arrotta_dexar_2022, gorski_explainable_2021}), or adopt Natural Language to increase intuitiveness and understandability of the explanation (e.g.,~\cite{arrotta_dexar_2022, le_grace_2020}). Textual explanation can also embed semantic features (e.g.,~\cite{hadash_improving_2022}) that aid users' sense-making process.

Additionally, processing textual format imposes more cognitive load than visual format, so the content of the explanations should be presented in a sequential order to avoid cognitive overload. When explanations are presented gradually, users are able to understand the explanations better. 

\section{Discussion}

\subsection{More attention and new methods are needed to understand user needs for explanations.}
Based on our analysis, we find that gathering user requirements before developing an XAI system is not a widespread practice. This finding provides further empirical evidence for why existing XAI research lack usability, practical interpretability, and efficacy for real users~\cite{abdul2018trends, doshi2017towards,miller2019explanation, zhu2018explainable}. If researchers are not clear about users' needs, their skills, and how they process technical explanations, it is not surprising that real users have difficulties using their XAI systems.

We hypothesize that there are several reasons for this. First, XAI has so far been primarily a technical research field. Its focus has been developing new algorithms that make black-box ML {\em interpretable}. Under this discipline, XAI systems are mostly seen as a prototype to showcase and test algorithmic feasibility. However, interpretability only means that the resulting ML models {\em can} potentially be interpreted. A significant amount of research and HCI design is needed to turn something interpretable into actual explanations that real users find understandable and useful. Second, we believe another reason is that established user research methods are insufficient to gather users' explanation needs. Explanations of AI are highly technical in nature, and there have not been a lot of existing examples for an average user to establish expectations. As a result, most users have problems articulating what their needs are, when it comes to XAI. In turn, even if a research team carries out time-consuming user research tasks, they may not find enough information to extract well-defined user requirements. This methodological challenge is echoed in all aspects of UX design for ML and AI products\cite{Yang2020}. We suggest that XAI researchers can use first person methods\cite{Lucero2019}, such as autoethnography\cite{cecchinato2017always,lucero2018living} and autobiographical design methods\cite{desjardins2018revealing,neustaedter2012autobiographical}, to build on their own knowledge. We also suggest more collaboration with HCI designers to provide heuristic evaluations~\cite{nielsen1990heuristic} as experts on user interface and user experience. 


\subsection{Interactivity is a key towards real explanability and actionable understanding.}
An increasing amount of recent evidence suggests that interactive explanations can be more user-friendly~\cite {liao_questioning_2020,Miller2017ExplainableSciences,wang_designing_2019}. However, less than half of the EIs in our surveyed publications are interactive. Among them, many offer limited interactivity such as letting users click through a pre-defined sequence of explanation content. There is also a lack of XAI work on scaffolding, user engagement, and other design methods to structure user interaction. This significantly limits XAI's ability to support more complex interactions around explanations, which may be essential for non-experts. 
We encourage XAI researchers and UX practitioners design interactive comparing interface to address this needs. A promising direction is to design the interface with matrix information architecture, where it supports interactivity and organize explanations' features in multiple dimensions. Recent work such as \cite{villareale2024GameXAI} offers a promising direction towards exploring new interaction formats of EIs through computer games.

\subsection{Evaluation of EI and XAI should reduce confounding factors.}
Our findings suggest that while researchers have adopted a wide variety of evaluation metrics, most publications evaluated the performance of the larger system where the EI is a component of. For example, Hjorth~\cite{hjorth_naturallanguageprocesing4all_2021} designed the XAI interface to help students learn about the policy views and political communications through a machine learning system adopting Natural Language. While XAI interface is an important component to help users communicate with the system, he does not evaluate the interface itself, but evaluate the whole system. This means that there are a lot of confounding factors when XAI researchers try to derive implications from their evaluation results. When the users report overall satisfaction with the entire system, it is difficult to tease out which components contribute to it and in what ways. We encourage future EI research to conduct more targeted evaluation of EIs and their components before evaluating the larger system where XAI is part of. 

Related, XAI researchers heavily rely on predefined desiderata (e.g., {\em trust, causality, transferability, informativeness, and fair and ethical decision making}~\cite{lipton2018mythos}) as evaluation metrics. While these desiderata offer a general direction, they are often too general to target the specific requirements or to be operationalized as evaluation metrics in a given context of use. For instance, what is considered trustworthy or informative can differ vastly from AI experts to novice end-users. Similarly, how to measure trust is a complicated task in itself. We believe that conducting user research, especially using qualitative methods (e.g., developing user group profiles, task models, and personas) at the early stage, can be a highly effective way to supplement the context-free desiderata.

\subsection{Reviewing framework as a generative tool}
The work here presented analyzes existing research utilizing eight properties. As seen in figures 2, 3, 5 and 6, the frequency distribution of publications per property is generally quite skewed, for example there is a lot of work that has sequential information structure while very few utilize hierarchical structure, or again the majority of publications examined adopts instructing interaction type while very few implemented a conversational approach. It is worth asking whether maybe there are opportunities delineated by the negative space shown by this analysis, for example there are no instances of EI that implement static interactivity AND hierarchical information architecture, is that because it simply does not make sense or could it be a missed opportunity for future work? 
Examining even just one of the clusters identified in section 4.4 it is possible to hypothesize EI that do not exist but might improve transparency and interpretability. For example looking at cluster 1, we could imagine an EI that utilizes UX and UI experts rather than domain experts and focuses on evaluation rather than analyses or production.  

The analysis here presented could be turned into a generative framework by looking at combinations of properties that are not represented in the dataset.

\section{Limitations}
There are several limitations of this paper. First, our systematic search is carried only in the ACM Digital Library database due to the HCI focus on our subject. It is likely that some relevant research is published in other venues, such as NeurIPS, ICML, or arXiv, and therefore is not included in our survey. Since {\em explainable interface} is an emerging terminology that XAI researcher only started using very recently, all papers must be manually scanned for relevance. A broader search is thus out of scope. We believe that, as one of the first survey on this topic, the number of papers resulted from our systematic search included in this paper is large enough to provide a representative sample of emerging research in explainable interface. Among the six related review articles, 50\% of them also only use the ACM Digital Library. 

Second, our analysis is based on what the researchers report in their publication. It is possible that, depending on the focus of the paper, researchers may omit certain details about user participation. For instance, if the researchers conducted informal user research to gather user needs, they may not report it in a publication where the focus is on XAI algorithms. Furthermore, there has been growing recognition of {\em first person methods}~\cite{Lucero2019} that uses HCI researchers' own experience for data collection, as opposed to external users. These first person methods are not considered as part of the process to define user requirements in our analysis.

\section{Conclusion}

Despite its technological breakthroughs, eXplainable Artificial Intelligence (XAI) research has limited success producing the {\em effective explanations} needed by users. In order to improve XAI systems' usability, practical interpretability, and efficacy for real users, the emerging area of {\em Explainable Interfaces} (EIs) focuses on the user interface and user experience design aspects of XAI. This paper presents a systematic survey of 53 publications to identify current trends in human-XAI interaction and promising directions for EI design and development. This is among the first systematic survey of EI research. 

In conclusion, we conducted a systematic literature review of 53 publications about the design and evaluation of EIs. Compared to existing technical review articles that focuses on {\em what} to explain, we focuses on {\em how} current XAI communicate the explanation to human users. In fact, is is only through unraveling the modalities of how EI can effectively function that we can radically improve future designs. Through three guiding research questions, we examined how participants are involved as well as how EIs are designed and evaluated. Using cluster analysis, we identify current trends in human-XAI interaction and promising directions for EI design and development. 

\begin{acks}
This work is supported by the Danish Novo Nordisk Foundation under Grant Number NNF20OC0066119. 
\end{acks}

\bibliographystyle{ACM-Reference-Format}
\bibliography{bibliography}

\appendix

\end{document}